\documentclass[aps,twocolumn,floatfix,showpacs,amssymb]{revtex4}
\usepackage{graphicx,bm}
\usepackage{hyperref}
\usepackage{amsmath}

\begin{document}
\title{Many Bosons in a Narrow Magnetic Feshbach Resonance}
\author{Ludovic Pricoupenko}
\affiliation
{
Laboratoire de Physique Th\'{e}orique de la Mati\`{e}re Condens\'{e}e, Universit\'{e} Pierre et Marie Curie and CNRS, 
4 place Jussieu, 75252 Paris, France
}
\date{\today}
\begin{abstract}
The many-boson problem in the presence of an asymptotically narrow Feshbach resonance is considered. 
The low energy properties are investigated using a two-channel Hamiltonian. The energy spectrum 
of this model is shown to be bounded from below in the limit of a zero range interaction. This implies the promising 
possibility of achieving a strongly interacting bosonic phase in a dilute regime where the details of the actual 
interatomic forces are irrelevant. The integral relation between the energy and the one-body momentum distribution 
is derived.
\end{abstract}
\pacs{67.85.-d,05.30.Jp,03.75.Nt}
%
\maketitle

The magnetic Feshbach resonance (FR) technique allows the study of highly correlated and dilute many-body systems 
in which the two-body scattering cross section is arbitrarily large. This has led to ways of studying the crossover 
between fermionic (BCS) and bosonic (BEC) superfluidity in the two spin-component Fermi gas \cite{Gio08} and to obtain 
the first evidence of Efimov states \cite{Efi70,Kra06,Kno09,Zac09,Gro09,Wen09,Wil09,Gro11,Ber11,Nak11,Mac12} as 
reviewed in Ref.~\cite{Fer10}. The large scale separation between the scattering length (denoted by $a$) and the 
interaction range (denoted by $b$, of the order of a few nanometers) suggests the existence of universal low energy 
properties in this system \cite{Ho04,Nav11,Rem12}. This universality for energies much smaller in absolute value 
than ${E_b = \hbar^2/(Mb^2)}$, where $M$ is the atomic mass, should be revealed using a model where the interaction 
range is arbitrarily small.
The prediction \cite{Ham07,Ste09,Del10,Del11a,Del11b} and the observation \cite{Fer09,Pol09} of universal four-body 
states are the first signatures of possible universal features for a many-boson system in the presence of three-body 
Efimov states.
Unfortunately, Efimov physics is also linked to large atomic losses observed in various experiments and
the stability issue of the degenerate resonant Bose gas is thus puzzling \cite{Rem12}. 
Eventually unlike the fermionic case, little is known about resonant many-boson systems which are now a fascinating 
issue at the cutting edge of the research in ultracold atoms.

Recently asymptotically narrow FRs have attracted large interest in the stability issue of the resonant degenerate 
gas \cite{Nis12}. As a consequence of the large scale separation between the Efimov ground state energy and the 
characteristic energy of deep bound states $E_b$, Efimov states have been predicted to be long lived 
\cite{Pet04b,Gog08,Lev09,Pri10b}. Moreover, three-body recombination towards deep molecules is also reduced 
\cite{Wan11}. Asymptotically narrow FRs are thus a relevant way to achieve a thermalized and dilute resonant 
many-boson system \cite{Ho12,Nis12}.
The FR mechanism is based on the coherent coupling of atomic pairs (belonging to an open channel) with a molecular 
state  (belonging to a closed channel). Using an external magnetic field ${\mathcal B}$ permits the tuning of the 
molecular binding energy ${E_{\rm mol}}$ with ${d E_{\rm mol}/d{\mathcal B}= \delta \mu}$. As a result, the 
scattering length in the vicinity of the FR located at ${\mathcal B=\mathcal B}_0$ can be fixed through the law 
${a = a_{\rm bg} [1-\Delta \mathcal B/(\mathcal B - \mathcal B_0)]}$ \cite{Moe95}. Here, ${\Delta {\mathcal B}}$ 
is the width of the resonance and ${a_{\rm bg}}$ is the background scattering length (it is assumed that 
${a_{\rm bg}}$ is of the order of $b$). 
For a weak coherent coupling between the atomic pairs and the molecular state, the resonance 
is narrow ${(|\delta \mu \Delta \mathcal B| \ll E_b)}$, and the width parameter
${R^\star=\hbar^2/(M a_{\rm bg} \delta \mu \Delta \mathcal B)}$ becomes essential in the description of low energy 
properties. Asymptotically narrow resonances where ${R^\star \gg b}$ and ${|a|\gg b}$, are characterized by the 
scattering amplitude \cite{Pet04b}
\begin{equation}
f_0(k) = - (1/a + R^\star k^2 + i k)^{-1} ,
\label{eq:scatt_amplitude}
\end{equation}
where the actual range of the interaction $b$ is negligible for ${k  \ll 1/\sqrt{b R^\star}}$.

In this Letter, a system of $N$ identical bosons is considered in an asymptotically narrow FR. It is shown that 
${(i)}$ the low energy properties of this system can be fully described by using the zero range limit of a 
two-channel model, ${(ii)}$ the ground state of this model corresponds to an actual strongly correlated long-lived 
and dilute bosonic phase, ${(iii)}$ at high momentum ${\mathbf k}$, the atomic momentum distribution ${n_{\mathbf k}}$  
behaves as
\begin{equation}
n_{\mathbf k} \underset{k\to \infty}{=} \frac{c_4}{k^4} + \frac{c_6}{k^6} + \dots \quad ,
\label{eq:tail}
\end{equation}
and ${(iv)}$ for temperatures much smaller than ${\hbar^2/(MbR^\star)}$, 
the mean energy $E$ of the system satisfies 
\begin{multline}
E - E^{\rm trap} = \frac{\hbar^2}{2M}\int \frac{d^3{k}}{(2\pi)^3} \left( k^2 n_{\mathbf k} -\frac{a^2 c_4}{1+k^2 a^2} \right) \\
+ \frac{\hbar^2 R^\star c_6}{8M\pi}-\langle \hat{K}^{\rm mol}\rangle ,
\label{eq:energy_theorem}
\end{multline}
where ${\langle \hat{K}^{\rm mol}\rangle}$ is the mean translational kinetic energy of the molecular state 
involved in the FR mechanism and ${E^{\rm trap}}$ is the contribution due to
the possible interaction with an external potential. Equation~\eqref{eq:energy_theorem} generalizes the Tan relation  
of Ref.~\cite{Tan08ab} for asymptotically narrow FRs.

In what follows, the system is described by using a two-channel model specific to narrow resonances. It is a simplified version of 
more general two-channel models used in the description of the quantum many-body problem \cite{Tom98,Tim99,Tim01,Hol01,Kok02}.
In the present two-channel model, the molecular state (of mass $2M$) is structureless: a reasonable approximation 
for the description of mechanisms at energies much smaller than ${E_b}$. The operator ${a_{\mathbf k}^\dagger}$ (or  
${b_{\mathbf k}^\dagger}$) creates an atomic (or a molecular) plane wave 
${\langle \mathbf r|{\mathbf k} \rangle=\exp({i {\mathbf k} \cdot {\mathbf r}})}$ and obeys 
the usual commutation rules for bosons. The kinetic energy of a particle of wave vector ${\mathbf k}$ is 
denoted by ${E_{\mathbf k} = \frac{\hbar^2 k^2}{2M}}$ so that in absence of an external potential, the free 
Hamiltonian for atoms and molecules is
\begin{equation}
H_0 =\int \frac{d^3k}{(2\pi)^3} \left[ E_{\mathbf{k}}  a^\dagger_{\mathbf k} a_{\mathbf k} 
+ \left( \frac{ E_{\mathbf k}}{2} +E_{\rm mol} \right) b^\dagger_{\mathbf k} b_{\mathbf k}
\right] .
\label{eq:H0}
\end{equation}
The interchannel coupling converts a molecule into a pair of atoms via the operator $V$ 
and vice-versa via the operator ${V^\dagger}$. It is modeled by a constant amplitude 
$\Lambda$ (chosen real positive) and a generic cutoff function 
${\chi_\epsilon(k)=\exp \left(-\frac{k^2 \epsilon^2}{4} \right)}$ (with ${\epsilon \ge 0}$)~,
\begin{equation}
V= \Lambda  \int 
\frac{d^3Kd^3k}{(2\pi)^6}  \, \chi_\epsilon(k) \, 
b_{\bf K} a^\dagger_{{{\mathbf K} }/{2}-{\mathbf k}} a^\dagger_{{{\mathbf K}}/{2}+{\mathbf k}}
\label{eq:V} .
\end{equation}
The Hamiltonian is thus ${H_0 + V + V^\dagger}$ and the scattering amplitude denoted by $f_\epsilon(k)$ 
at energy ${E=E_{\mathbf k}}$ is given by
\begin{equation}
\frac{M[\chi_\epsilon(k)]^{-2}}{4\pi \hbar^2f_\epsilon(k)}
= \frac{E_{\rm mol}-E}{2\Lambda^2} + \int \frac{d^3k'}{(2\pi)^3} \frac{[\chi_\epsilon(k')]^2}{z-2 E_{\mathbf k'}} , 
\label{eq:scatt_amplitude_2channel}
\end{equation}
where the variable $z$ encapsulates the usual prescription ${z=E+i0^+}$ for ${E>0}$ and ${z=E}$ otherwise. For 
vanishing values of $\epsilon$, the identification of $f_\epsilon(k_0)$ with Eq.~\eqref{eq:scatt_amplitude} gives
\begin{equation}
E_{\rm mol} = 
\frac{\hbar^2}{M R^\star} \left( \frac{\sqrt{2}}{\epsilon \sqrt{\pi}} - \frac{1}{a}\right)
\quad \mbox{and} \quad 
\Lambda = \frac{\hbar^2}{M} \sqrt{\frac{2\pi}{R^\star}} .
\label{eq:parameters}
\end{equation}
In what follows, only states with asymptotically small values of $\epsilon$ are considered while ${a}$ 
and ${R^\star}$ are kept fixed. In this formal zero range limit, Eq.~\eqref{eq:parameters} shows that 
${E_{\rm mol}}$ has an arbitrarily large positive value. Thus, the molecular state is occupied only 
through virtual processes. This modeling is valid in the vicinity of the resonance, i.e., 
${|\mathcal B-\mathcal B_0|\ll |\Delta \mathcal B|}$, otherwise off-resonant effects must be taken 
into account \cite{Pri11c}. In this small detuning regime the actual molecular energy is of the order 
of ${\delta \mu \Delta B}$ (see, for instance, Eqs.(19), (21) and (22) in Ref.~\cite{Pri11c}) which is 
a high energy scale with respect to the natural energy cutoff in two-body processes, i.e., 
${\hbar^2/(MR^\star\, ^2)}$. Comparing ${E_{\rm mol}}$ in Eq.~\eqref{eq:parameters} with 
${{\hbar^2}/{(M R^\star a_{\rm bg})}}$, one finds that the short range parameter ${\epsilon}$ plays the 
role of a potential radius, which is as expected.

As a consequence of the interchannel coupling, an eigenstate for a system composed of $N$ atoms is in 
general a coherent superposition of states containing ${0}$ to ${[N/2]}$ molecules
\begin{equation}
|\Psi_\epsilon \rangle = \sum_{m=0}^{[N/2]} |\psi_\epsilon^m \rangle ,
\end{equation}
where ${|\psi_\epsilon^m \rangle}$ is a state with ${m}$ molecules and ${N-2m}$ atoms. Projection of the 
stationary Schr\"{o}dinger's equation at energy $E$ in the subspace with $m$ molecules (${1 < m < [N/2]}$) 
yields the recursive equation
\begin{equation}
(E-H_0) |\psi_\epsilon^m \rangle = V |\psi_\epsilon^{m+1} \rangle + V^\dagger |\psi_\epsilon^{m-1} \rangle . 
\label{eq:recursion}
\end{equation}
The recursion begins for ${m=0}$ by the equation
\begin{equation}
|\psi_\epsilon^0 \rangle = |\Phi_{\rm inc} \rangle +  G_0(z) V |\psi_\epsilon^1 \rangle ,
\label{eq:recursive_0}
\end{equation}
where ${G_0(z) = (z-H_0)^{-1}}$ is the resolvent and ${|\Phi_{\rm inc} \rangle}$ is the possible incoming wave 
for a scattering process when ${E \ge 0}$, whereas ${|\Phi_{\rm inc} \rangle\equiv 0}$ when ${E<0}$. The recursion 
for ${m=1}$ plays a central role in the rest of this Letter. From Eqs.~\eqref{eq:recursion} and \eqref{eq:recursive_0} 
it takes the form
\begin{equation}
[ E-H_0 - V^\dagger G_0(z)V ] |\psi_\epsilon^1 \rangle = V^\dagger |\Phi_{\rm inc} \rangle + V |\psi_\epsilon^2 \rangle.
\label{eq:formal_1mol}
\end{equation}
In the subsequent lines, the momentum representation is used and 
${(k)_p = (\mathbf k_1,\mathbf k_2 \dots ,\mathbf k_p)}$ [or $(K)_m = (\mathbf K_1,\mathbf K_2 \dots ,\mathbf K_m)$] 
denotes the wave vectors of $p$ particles (or of $m$ molecules). The $m$-molecule state can then be written as
\begin{multline}
|\psi_\epsilon^{m}\rangle = \int \frac{d(k)_p d(K)_m}{(2\pi)^{3p+3m}}  
\frac{\langle (k)_p ; (K)_m | \psi_\epsilon^{m} \rangle}{\sqrt{p!}\sqrt{m!}}\\
a_{\mathbf k_1}^\dagger \dots a_{\mathbf k_p}^\dagger b_{\mathbf K_1}^\dagger \dots b_{\mathbf K_m}^\dagger |0\rangle .
\end{multline}
In the following, a superscript ${\hat{i}}$ in ${(k)_p^{\hat{i}}}$ means that the wave vector ${\mathbf k_i}$ is withdrawn 
from ${(k)_p}$. For instance, ${(k)_N^{\hat{1}\hat{2}}=(\mathbf k_3, \mathbf k_4, \dots \mathbf k_N)}$. The ket 
${|(k)_p ; (K)_m \rangle}$ describes the state where the atom $i$ (${1\le i\le p}$) is in the plane wave of wave vector 
${\mathbf k_i}$, and the molecule $j$ (${1\le j \le m}$) is in the plane wave of momentum ${\mathbf K_j}$. Furthermore 
${\mathbf k_{ij}=\frac{\mathbf k_i-\mathbf k_j}{2}}$ (or ${\mathbf K_{ij}=\mathbf k_i+\mathbf k_j}$) denotes the relative 
(or the total) wave vector for the pair ${(i,j)}$. 

As in the two-body states, for an arbitrarily large relative wave vector of the pair ${(ij)}$, it is expected that in the 
presence of ${m+1}$ molecules (i.e., ${|\psi_\epsilon^{m+1}\rangle \ne 0}$) ${\langle (k)_p;(K)_{m}|\psi_\epsilon^m \rangle}$ 
has the asymptotic  behavior ${\chi_\epsilon(k_{ij})/k_{ij}^2}$ \cite{UVlimit}. This behavior is obtained  by considering 
Eq.~\eqref{eq:recursion} divided by ${E_{\rm mol}\equiv O(1/\epsilon)}$ in the zero range limit; the only nonvanishing 
contribution in the right-hand side of the resulting equation is ${\langle (k)_p ; (K)_m | V^\dagger | \psi_\epsilon^{m-1} \rangle}$. 
This enables the ${m}$-molecule wave function to be expressed in terms of the ${(m-1)}$-molecule wave function:
\begin{multline}
\lim_{\epsilon \to 0} \langle (k)_{p};(K)_m|\psi_\epsilon^{m} \rangle= 
-\sqrt{\frac{(p+2)!R^\star}{(2 m)^3 \pi p! }} \sum_{i=1}^m \lim_{k\to  \infty} \bigg[ k^2 
\\
\times \lim_{\epsilon \to 0} 
\langle (k)_{p}, \frac{\mathbf K_i}{2}-\mathbf k, \frac{\mathbf K_i}{2}+\mathbf k;(K)_m^{\hat{i}}|\psi_\epsilon^{m-1} \rangle \bigg] .
\label{eq:un_sur_k2}
\end{multline}
In the center-of -mass frame, ${|\psi_\epsilon^{1} \rangle}$ can be factorized by a ket of ${N-2}$ particles, which is  
denoted by ${|D\rangle}$ in the zero range limit: 
\begin{equation}
\frac{\displaystyle \lim_{\epsilon \to 0^+} \langle (k)_N^{\hat{1}\hat{2}} ; \mathbf K_{12} |\psi_\epsilon^{1} \rangle}
{(2\pi)^3 \sqrt{R^\star N(N-1)}}
 = \delta(\sum_{i=3}^N \mathbf k_i + \mathbf K_{12}) 
\langle (k)_N^{\hat{1}\hat{2}} | D \rangle .
\end{equation}
Similarly, the possible incoming state is factorized as
${
\langle (k)_N |\Phi_{\rm inc} \rangle = (2\pi)^3 \delta(\sum_{i=1}^N \mathbf k_i ) \langle (k)_N | \phi_{\rm inc} \rangle
}$. 
The equation for ${|D\rangle}$ is deduced from Eqs.~\eqref{eq:V},\eqref{eq:formal_1mol} and \eqref{eq:un_sur_k2}
and the identity $\lim_{k_{ij} \to \infty } k_{ij}^2 \langle (k)_N^{\hat{1}\hat{2}}| D \rangle = 
\lim_{k_{12} \to \infty } k_{12}^2 \langle (k)_N^{\hat{i}\hat{j}}| D \rangle$ which comes from the exchange symmetry 
of the two molecules described by ${|\psi^2_\epsilon\rangle}$ for ${\{i,j\}\notin\{1,2\}}$ \cite{unexpected}:
\begin{multline}
\int \frac{d^3k_2}{(2\pi)^3} \frac{ \operatornamewithlimits{\sum}_{3\le i<j\le N} \langle (k)_N^{\hat{i}\hat{j}}|D \rangle 
+ 2 \sum_{i=3}^N \langle (k)_N^{\hat{1}\hat{i}}|D \rangle}
{\sum_{i=2}^N k_i^2 + \sum_{2\le i<j\le N} \mathbf k_i \cdot \mathbf k_j - M z/\hbar^2} \\
+ \frac{R^\star}{4\pi} \sum_{3\le i<j\le N} 
\lim_{k_{12} \to \infty } \left[ k_{12}^2 \langle (k)_N^{\hat{i}\hat{j}}| D\rangle \right] \\
- \int  \frac{d^3k_{12}}{(2\pi)^3} \langle (k)_N | \phi_{\rm inc} \rangle  = \frac{\langle (k)_N^{\hat{1}\hat{2}}| D\rangle }
{4\pi f_0(k^{\rm col})} , 
\label{eq:STM_many_bosons}
\end{multline}
where ${k^{\rm col}=\sqrt{M E^{\rm col}}/\hbar}$ is the collisional wave number for the pair ${(1,2)}$ defined from the
energy ${E^{\rm col} = E - \sum_{n=3}^N E_{\mathbf k_n}-E_{\mathbf K_{12}}/2}$. 
Surprisingly Eq.~\eqref{eq:STM_many_bosons} supports solutions where 
${\lim_{\epsilon \to 0} |\psi_\epsilon^{m} \rangle=0}$ for ${m \ge 2}$ and thus 
${\lim_{k_{ij} \to \infty } k_{ij}^2 \langle (k)_N^{\hat{1}\hat{2}}| D\rangle = 0}$ for ${\{i,j\}\notin\{1,2\}}$ 
\cite{one-mol-state}.

Without loss of generality, the energy spectrum of Eq.~\eqref{eq:STM_many_bosons} is shown to be bounded from below in the 
strict resonant limit ${|a|=\infty}$. To show this, we use the following inequality verified by negative energy solutions 
of Eq.~\eqref{eq:STM_many_bosons}:
\begin{multline}
q R^\star |\langle (k)_N^{\hat{1}\hat{2}}|D\rangle|  < \frac{R^\star}{4\pi q} \sum_{3\le i<j\le N} 
\lim_{k_{12} \to \infty } \left| k_{12}^2 \langle (k)_N^{\hat{i}\hat{j}}| D\rangle \right|\\
 + \sum_{i<j ; (i,j)\ne (1,2)} 
\int \frac{d^3k_{12}}{2\pi^2q^3} \frac{|\langle (k)_N^{\hat{i}\hat{j}}|D\rangle|}{k_{12}^2/q^2+1} ,
\label{eq:STM_qlarge}
\end{multline}
where ${q={\sqrt{-M E}}/{\hbar}}$ is the binding wave number. Then, we consider a hypothetical solution of Eq.~\eqref{eq:STM_many_bosons} 
having an arbitrarily large and negative energy (${q\to \infty}$). Assuming that ${\langle (k)_N^{\hat{i}\hat{j}}|D\rangle}$ is a 
bounded function, integration in the domain where ${k_{12}}$ is smaller or of the order of ${q}$ in the right-hand side of 
Eq.~\eqref{eq:STM_qlarge} gives a finite result in the limit where ${q}$ is arbitrarily large. In the domain where ${k_{12}}$ 
is much larger than ${q}$, the high momentum behavior of ${\langle (k)_N^{\hat{i}\hat{j}}|D\rangle}$ \cite{one-mol-state} 
gives rise to a vanishing contribution after integration in Eq.~\eqref{eq:STM_qlarge}. As for any fixed value of 
${(k)_N^{\hat{1}\hat{2}}}$, the left-hand side of Eq.~\eqref{eq:STM_qlarge} is bounded from above by a finite quantity, 
${q R^\star}$ cannot be arbitrarily large. The desired result is thus obtained by contradiction.

From now on the effective range model (ERM) is used to derive Eqs.~\eqref{eq:tail} and \eqref{eq:energy_theorem}. 
To avoid any confusion with the two-channel model, the eigenstates in the ERM are underlined (as 
${|\underline{\Psi}\rangle}$, for instance). The ERM is a one-channel and zero range approach which supports the 
scattering amplitude in Eq.~\eqref{eq:scatt_amplitude} whereas the molecular field is hidden. 
The main idea of the ERM is to filter asymptotically (i.e.,  at interparticle distances much larger than $b$) 
correct wave functions among the set of singular wave functions satisfying a Schr\"{o}dinger's equation for 
pointlike interactions. This filtering process can be performed by using a pseudopotential. Following 
Refs.~\cite{Pri08,Pri11a}, the pseudopotential of the ERM is defined by introducing a Gaussian shape 
${\langle \mathbf r |\delta_\epsilon \rangle}$ tending to the Dirac distribution in the limit where 
the short range parameter ${\epsilon}$ goes to zero. In the momentum representation 
${\langle {\mathbf k}|\delta_\epsilon \rangle = \chi_\epsilon(k)}$; i.e., it coincides with the interchannel 
coupling function of the two-channel model. For a many-body system, a short range parameter ${\epsilon_{ij}}$ 
and a pseudopotential ${V_{ij}^{\epsilon_{ij}}}$ are introduced for each interacting pair of particles ${(i,j)}$:
\begin{equation}
V_{ij}^{\epsilon_{ij}} \, \cdot \,  = \frac{4\pi\hbar^2a}{M} |\delta_{\epsilon_{ij}} \rangle \lim_{{\epsilon_{ij}} \to 0^+}  
\left(\partial_{\epsilon_{ij}} - \sqrt{\frac{\pi}{2}} \frac{R^\star}{2} \partial_{\epsilon_{ij}}^2  \right) 
\epsilon_{ij} \langle \delta_{\epsilon_{ij}} | \, \cdot \, 
\label{eq:VLambda_epsilon}
\end{equation}
where the ket ${|\delta_{\epsilon_{ij}} \rangle}$ is associated with the function ${\delta_{\epsilon_{ij}}}$ 
for the relative coordinates of the pair. Instead of introducing the pseudopotential, the filtering process 
of the ERM can be defined through the notion of contact condition. To see this, one can notice that the action 
of the pseudopotential ${V_{ij}^{\epsilon_{ij}}}$  on an eigenstate ${|\underline{\Psi}\rangle}$ gives rise to 
a $\delta$-source term for each interacting pair in the stationary Schr\"{o}dinger's equation
\begin{equation}
\left( E- \sum_{i=1}^{N} \frac{\hbar^2 \hat{\mathbf k}_i^2}{2M}\right) |\underline{\Psi} \rangle 
= \frac{4\pi\hbar^2}{M}\sum_{i<j} |A_\Psi^{(ij)}, \delta_{\epsilon_{ij}} \rangle.
\label{eq:Schrodi_N}
\end{equation}
Here, ${|A_\Psi^{(ij)}\rangle}$ depends on the set of short range parameters ${\{\epsilon_{pq}\}}$ 
such that ${(p,q)\ne(i,j)}$ and verifies
${
|A^{ij}_\Psi \rangle=-\sqrt{\pi/2} \lim_{\epsilon_{ij}\to 0^+} 
\left( \epsilon_{ij} \langle \delta_{\epsilon_{ij}} | \underline{\Psi} \rangle \right)
}$ \cite{Identity}. The ERM can then be defined from Eq.~\eqref{eq:Schrodi_N} together 
with a boundary condition at the contact of each interacting pair ${(i,j)}$ \cite{Identity}
\begin{equation}
\lim_{\epsilon_{ij} \to 0^+} B_{\epsilon_{ij}} \langle \delta_{\epsilon_{ij}} | \underline{\Psi} \rangle = 0 ,
\label{eq:boundary_condition}
\end{equation}
where $B_\epsilon$ is the boundary operator defined by
\begin{equation}
B_\epsilon \, \cdot \, =
\left[\partial_\epsilon - \sqrt{\frac{\pi}{2}} \frac{R^\star}{2} \partial_\epsilon^2 + \sqrt{\frac{\pi}{2}} \frac{1}{a} \right] 
\epsilon  \, \cdot  \, \quad .
\label{eq:Boundary_operator_epsilon}
\end{equation}

From Eq.~\eqref{eq:Schrodi_N} one finds that two nondegenerate eigenstates of the ERM, say, 
${|\underline{\Psi} \rangle}$ (of energy ${E_\Psi}$) and ${|\underline{\Phi}\rangle}$ (of 
energy ${E_\Phi \ne E_\Psi}$), are not mutually orthogonal:
\begin{equation}
\langle \underline{\Phi} |\underline{\Psi} \rangle 
= \frac{4\pi\hbar^2}{M}\sum_{i<j} \frac{\langle \underline{\Phi} |A_\Psi^{(ij)},\delta_{\epsilon_{ij}}\rangle
- \langle A_\Phi^{(ij)}, \delta_{\epsilon_{ij}} |\underline{\Psi} \rangle}{E_\psi-E_\Phi} .
\label{eq:prod_scal}
\end{equation}
Interestingly, a modified scalar product denoted by ${(\cdot |\cdot)_0}$  which makes the ERM self-adjoint
can be constructed by using Eqs.~\eqref{eq:boundary_condition},\eqref{eq:Boundary_operator_epsilon}  and \eqref{eq:prod_scal},
\begin{equation}
(\underline{\Phi}|\underline{\Psi})_0 = \lim_{\{\epsilon\} \to 0}
\biggl(\,
\prod_{i<j} B_{\epsilon_{ij}} 
\biggr)
\langle \underline{\Phi} |\underline{\Psi} \rangle ,
\label{eq:modified_scalar}
\end{equation}
where ${\lim_{\{\epsilon\}\to 0}}$ denotes the limit in which all the short range parameters $(\epsilon_{ij})$ 
tend to zero. Equation \eqref{eq:modified_scalar} generalizes the modified scalar product introduced in Ref.~\cite{Pri06a} 
for two interacting particles. The equivalence of this modified scalar product with the usual scalar product of the two-channel 
model has been already depicted at the two-body level in Ref.~\cite{Pri11d}. This mapping can be generalized in the $N$-body case 
as follows. In the zero range limit, analogously to Eq.~\eqref{eq:un_sur_k2}, the hidden ${m}$-molecule wave function is revealed 
in the state ${|\underline{\Psi}\rangle}$ by the existence of a ${k_{ij}^{-2}}$ behavior in the large momentum limit for ${m}$ 
distinct pairs ${(ij)}$. To be more explicit, one introduces the state ${| A^{(ij),\dots (ln),(pq)}_\Psi \rangle}$ \cite{independent} 
corresponding to the contact of the two particles in the distinct pairs ${(ij)}$,\dots ${(ln),(pq)}$. It is defined iteratively 
from the state ${| A^{(ij)}\rangle}$ by 
\begin{multline}
\langle (k)_N^{\hat{i}\hat{j}\dots \hat{l}\hat{n}}; \mathbf K_{ij} \dots \mathbf K_{ln} | A^{(ij),\dots (ln)}_\Psi \rangle
\underset{k_{pq}\to \infty}{=} 
\frac{\chi_{\epsilon_{pq}}(k_{pq})}{k_{pq}^2}\\
\times 
\langle (k)_N^{\hat{i}\hat{j}\dots \hat{l}\hat{n}\hat{p}\hat{q}}; \mathbf K_{ij} \dots \mathbf K_{ln} | A^{(ij),\dots (ln),(pq)}_\Psi \rangle .
\end{multline}
In the scalar product ${\langle \underline{\Phi} |\underline{\Psi} \rangle}$, the contact for $m$ distinct pairs ${(ij),\dots (ln)}$ 
manifests itself in the limit where ${\{\epsilon\} \to 0}$ by the term
\begin{multline}
(-1)^{m}\frac{16\pi^2 }{(2\pi)^{3m/2}}
\epsilon_{ij} \dots \epsilon_{ln} \langle A^{(ij),\dots (ln)}_\Phi | A^{(ij),\dots (ln)}_\Psi \rangle .
\label{eq:monom}
\end{multline}
The number of terms of the same type as in Eq.~\eqref{eq:monom} is equal to ${P_m^N=N!/[2^m m!(N-2m)!]}$ i.e., the number of 
collections of $m$ distinct pairs taken from the set of $N$ particles. Applying all of the boundary operators to the right 
hand side of Eq.~\eqref{eq:modified_scalar} and identifying the result with the scalar product in the two-channel model, 
one finds the following mapping in the zero range limit:
\begin{equation}
\lim_{\{\epsilon\} \to 0}|A^{(ij),\dots (ln)}_\Psi \rangle
= \left(\frac{4\pi}{R^\star}\right)^{m/2} \frac{1}{4\pi \sqrt{P_m^N}} \lim_{\epsilon\to 0} |\psi^m_\epsilon \rangle
\label{eq:correspondance}
\end{equation}
and ${\lim_{\{\epsilon \}\to 0}|\underline{\Psi}\rangle = \lim_{\epsilon \to 0^+} |\psi^0_{\epsilon}\rangle}$. 
Equation~\eqref{eq:STM_many_bosons} is recovered by using this mapping and applying the contact condition in 
Eq.~\eqref{eq:boundary_condition} on the state ${|\underline{\Psi}\rangle}$. 
For an eigenstate normalized with Eq.~\eqref{eq:modified_scalar}, i.e., ${(\underline{\Psi} |\underline{\Psi})_0=1}$, 
the one-particle momentum distribution (taking into account the contributions of the ${m}$-molecule states) is 
${n_{\mathbf k} =  ( \underline{\Psi} | a_{\mathbf k}^\dagger a_{\mathbf k} |\underline{\Psi})_0}$. This definition yields the correct 
normalization, i.e., 
\begin{equation}
\int \frac{d^3k}{(2\pi)^3} n_{\mathbf k} =N-2N^{\rm mol} \ 
; \ N^{\rm mol} \underset{\epsilon \to 0}{=} \sum_{m=1}^{[N/2]} m \langle \psi^m_\epsilon |\psi^m_\epsilon \rangle
\end{equation}
where ${N^{\rm mol}}$ is the mean number of molecules \cite{Identity2}. The ERM gives 
${c_4= {8 \pi N^{\rm mol}}/{R^\star}}$ in Eq.~\eqref{eq:tail}: a result which also can be 
deduced from the Hellmann-Feynman theorem~\cite{Wer09,Wer12ab}. The expression of the
coefficient ${c_6}$ in Eq.~\eqref{eq:tail} depends on the possible interaction of the 
particles with a trapping potential ${V^{\rm trap}(\mathbf r)}$: 
\begin{multline}
c_6 = \frac{32\pi^2 M}{\hbar^2} N(N-1) 
( A_\Psi^{(12)} | E - 2 {V}^{\rm trap}(\mathbf R_{12}) \\
- \sum_{n=3}^N \left[ \frac{\hbar^2 \hat{\mathbf k}_n^2}{2M} + V^{\rm trap}(\mathbf r_n)\right] |A_\Psi^{(12)})_0 .
\label{eq:c6}
\end{multline}
Equation~\eqref{eq:energy_theorem} is obtained using the identity  
${( \underline{\Psi} | V_{ij}^{\epsilon_{ij}} |\underline{\Psi})_0 = 0 
}$, which is valid provided that the state ${|\underline{\Psi} \rangle}$ verifies the 
contact condition in Eq.~\eqref{eq:boundary_condition} for all pairs ${(i,j)}$. Thus, applying 
${\lim_{\{\epsilon\} \to 0} \biggl(\prod_{i<j} B_{\epsilon_{ij}} \biggr)}$ on the mean value 
of the Hamiltonian of the ERM evaluated for the normalized eigenstate ${|\underline{\Psi}\rangle}$ 
yields
\begin{equation}
E - E^{\rm trap} = (\underline{\Psi} | \int \frac{d^3k}{(2\pi)^3} 
\frac{\hbar^2 k^2}{2M}  a_{\mathbf k}^\dagger a_{\mathbf k} |\underline{\Psi} )_0 ,
\label{eq:E_Th}
\end{equation}
where ${E^{\rm trap} = ( \underline{\Psi} | 
\sum_{n=1}^N V^{\rm trap}(\mathbf r_n) |\underline{\Psi} )_0}$ includes the contribution 
of the molecular state. The regularization of the kinetic energy in the right-hand side of 
Eq.~\eqref{eq:E_Th} through the modified scalar product permits to recover 
Eq.~\eqref{eq:energy_theorem} with ${\langle \hat{K}^{\rm mol}\rangle= \langle \Psi| \int \frac{d^3k}{(2\pi)^3} 
b^\dagger_{\mathbf k} b_{\mathbf k} E_{\mathbf k}|\Psi \rangle/2}$. 
Equation~\eqref{eq:energy_theorem} is 
valid for any linear combination of eigenstates of the ERM and thus also applies for low temperatures 
 ${k_BT \ll |\delta\mu \Delta B|}$.

In this many-boson problem, $R^\star$ plays the key role of natural short length cutoff and ensures 
the existence of a ground state in the formal zero range limit. This supports the existence of an 
actual long-lived dilute resonant bosonic phase which is denoted as the $R^\star$ phase. The mapping 
with the ERM permits us to obtain the integral relation satisfied by the mean energy and the one-particle 
momentum distribution. In the ERM, the hidden molecular field is described through singularities in 
the $N$-particle wave function at the contact of each interacting pair. The shallowness of the Efimov 
ground state is at the heart of the stability issue of the $R^\star$ phase. For three resonant bosons, 
the atoms experience the standard attractive ${1/R^2}$ effective potential for large values of the 
hyper-radius $R$ (${R \ll |a|}$) whereas the natural short length cutoff ${R^\star}$ linked to the 
width of the FR imposes a molecule-atom distance larger than $R^\star$ \cite{Gog08,Wan11}. One thus 
expects that the ground state is organized in pairs of bosons with a large population of the molecular 
state and a mean intermolecular distance of the order of $R^\star$. The liquid or crystalline nature 
of the ${R^\star}$ phase is an open fascinating issue. Narrow FRs have been found in many atomic species 
\cite{Chi10} and new techniques allowing a fine-tuning of the magnetic field are promising for the 
observation of the ${R^\star}$ phase \cite{Pas12}.

Stimulating discussions with Yvan Castin, Tom Kristensen, Dima Petrov, Gora Shlyapnikov and F\'{e}lix 
Werner are warmly acknowledged.


\begin{thebibliography}{0}
\expandafter\ifx\csname natexlab\endcsname\relax\def\natexlab#1{#1}\fi
\expandafter\ifx\csname bibnamefont\endcsname\relax
  \def\bibnamefont#1{#1}\fi
\expandafter\ifx\csname bibfnamefont\endcsname\relax
  \def\bibfnamefont#1{#1}\fi
\expandafter\ifx\csname citenamefont\endcsname\relax
  \def\citenamefont#1{#1}\fi
\expandafter\ifx\csname url\endcsname\relax
  \def\url#1{\texttt{#1}}\fi
\expandafter\ifx\csname urlprefix\endcsname\relax\def\urlprefix{URL }\fi
\providecommand{\bibinfo}[2]{#2}
\providecommand{\eprint}[2][]{\url{#2}}

\end{thebibliography}


\begin{thebibliography}{99}

\bibitem{Gio08} S. Giorgini, L.P. Pitaevskii, and S. Stringari, 
\href{http://dx.doi.org/10.1103/RevModPhys.80.1215}
{Rev. Mod. Phys. {\bf 80}, 1215 (2008).}

\bibitem{Efi70} V. Efimov, 
\href{http://linkinghub.elsevier.com/retrieve/pii/0370269370903497}
{Phys. Lett. {\bf 33}B, 563 (1970).}

\bibitem{Kra06} T. Kraemer {\sl et al.}, 
\href{http://www.nature.com/nature/journal/v440/n7082/abs/nature04626.html}
{Nature (London) {\bf 440}, 315 (2006).}

\bibitem{Kno09} S. Knoop {\sl et al.}, 
\href{http://www.nature.com/nphys/journal/v5/n3/abs/nphys1203.html}
{Nature Phys. {\bf 5}, 227 (2009).}

\bibitem{Zac09} M. Zaccanti {\sl et al.}, 
\href{http://www.nature.com/nphys/journal/v5/n8/abs/nphys1334.html}
{Nature Phys. {\bf 5}, 586 (2009).}

\bibitem{Gro09} N.~Gross, Z.~Shotan, S.~Kokkelmans, and L.~Khaykovich, 
\href{http://link.aps.org/doi/10.1103/PhysRevLett.103.163202}
{Phys. Rev. Lett. {\bf 103}, 163202 (2009).}

\bibitem{Wen09} A. N. Wenz {\sl et al.}, 
\href{http://link.aps.org/doi/10.1103/PhysRevA.80.040702}
{Phys. Rev. A {\bf 80}, 040702(R) (2009).}

\bibitem{Wil09} J.R. Williams {\sl et al.}, 
\href{http://link.aps.org/doi/10.1103/PhysRevLett.103.130404}
{Phys. Rev. Lett. {\bf 103}, 130404 (2009).}

\bibitem{Gro11} N. Gross {\sl et al.}, 
\href{http://dx.doi.org/10.1016/j.crhy.2010.10.004}
{C. R. Physique {\bf 12}, 4 (2011).}

\bibitem{Ber11} M. Berninger {\sl et al.}, 
\href{http://link.aps.org/doi/10.1103/PhysRevLett.107.120401}
{Phys. Rev. Lett. {\bf 107}, 120401 (2011).}

\bibitem{Nak11} S. Nakajima, M.~Horikoshi, T.~Mukaiyama, P.~Naidon, M.~Ueda, 
\href{http://link.aps.org/doi/10.1103/PhysRevLett.106.143201}
{Phys. Rev. Lett. {\bf 106}, 143201 (2011).}

\bibitem{Mac12} O. Machtey, Z. Shotan, N. Gross, L. Khaykovich 
\href{http://link.aps.org/doi/10.1103/PhysRevLett.108.210406}
{Phys. Rev. Lett. {\bf 108}, 210406 (2012).}

\bibitem{Fer10} F. Ferlaino and R. Grimm, 
\href{http://dx.doi.org/10.1103/Physics.3.9}
{Physics {\bf 3}, 9 (2010).} 

\bibitem{Ho04} T.-L. Ho, 
\href{http://dx.doi.org/10.1103/PhysRevLett.92.090402}
{Phys. Rev. Lett. {\bf 92}, 090402 (2004).}

\bibitem{Nav11} N. Navon {\sl et al.}, 
\href{http://dx.doi.org/10.1103/PhysRevLett.107.135301}
{Phys. Rev. Lett. {\bf 107}, 135301 (2011).}

\bibitem{Rem12} B.S. Rem {\sl et al.}, 
\href{http://arxiv.org/pdf/1212.5274}
{arXiv:1212.5274 .}

\bibitem{Ham07} H.-W. Hammer and L. Platter, 
\href{http://epja.edpsciences.org/articles/epja/abs/2007/05/10050_2007_Article_100303/10050_2007_Article_100303.html}
{Eur. Phys. J. A {\bf 32}, 113 (2007).}

\bibitem{Ste09} J. von Stecher, J.P. D'Incao, C.H. Greene, 
\href{http://www.nature.com/nphys/journal/v5/n6/abs/nphys1253.html}
{Nature Phys. {\bf 5}, 417 (2009).}

\bibitem{Del10} A. Deltuva,
\href{http://link.aps.org/doi/10.1103/PhysRevA.82.040701}
{Phys. Rev. A {\bf 82}, 040701(R) (2010).}

\bibitem{Del11a} A. Deltuva, 
\href{http://dx.doi.org/10.1209/0295-5075/95/43002}
{Eur. Phys. Lett. {\bf 95}, 43002 (2011).}

\bibitem{Del11b} A. Deltuva, 
\href{http://link.aps.org/doi/10.1103/PhysRevA.84.022703}
{Phys. Rev. A {\bf 84}, 022703 (2011).}

\bibitem{Fer09} F. Ferlaino {\sl et al.}, 
\href{http://link.aps.org/doi/10.1103/PhysRevLett.102.140401}
{Phys. Rev. Lett. {\bf 102}, 140401 (2009).}

\bibitem{Pol09} S.E. Pollack, D. Dries, and R. G. Hulet, 
\href{http://www.sciencemag.org/cgi/content/abstract/sci;326/5960/1683}
{Science {\bf 326}, 1683 (2009).}

\bibitem{Nis12} Y. Nishida, 
\href{http://dx.doi.org/10.1103/PhysRevLett.109.240401}
{Phys. Rev. Lett. {\bf 109}, 240401 (2012).}

\bibitem{Pet04b} D.S. Petrov, 
\href{http://dx.doi.org/10.1103/PhysRevLett.93.143201}
{Phys. Rev. Lett. {\bf 93}, 143201 (2004).}

\bibitem{Gog08} A.O. Gogolin, C. Mora, and R. Egger, 
\href{http://link.aps.org/doi/10.1103/PhysRevLett.100.140404}
{Phys. Rev. Lett. {\bf 100}, 140404 (2008).}

\bibitem{Lev09} J. Levinsen, T.G. Tiecke, J.T.M. Walraven, and D.S. Petrov, 
\href{http://link.aps.org/doi/10.1103/PhysRevLett.103.153202}
{Phys. Rev. Lett. {\bf 103}, 153202 (2009).}

\bibitem{Pri10b} L. Pricoupenko, 
\href{http://dx.doi.org/10.1103/PhysRevA.82.043633}
{Phys. Rev. A {\bf 82}, 043633 (2010).}

\bibitem{Wan11} Y. Wang, J. P. D'Incao, and B. D. Esry, 
\href{http://dx.doi.org/10.1103/PhysRevA.83.042710}
{Phys. Rev. A {\bf 83}, 042710 (2011).}

\bibitem{Ho12} T.-L Ho, X. Cui, and W. Li, 
\href{http://dx.doi.org/10.1103/PhysRevLett.108.250401}
{Phys. Rev. Lett. {\bf 108}, 250401 (2012).}

\bibitem{Moe95} {A. J. Moerdijk, B. J. Verhaar, and A. Axelsson, }
\href{http://link.aps.org/doi/10.1103/PhysRevA.51.4852}
{Phys. Rev. A {\bf 51}, 4852 (1995).}

\bibitem{Tan08ab} S. Tan, 
\href{http://dx.doi.org/10.1016/j.aop.2008.03.004}
{Ann. Phys. (N.Y.) {\bf 323}, 2952 (2008).}

\bibitem{Tom98} P. Tommasini {\sl et al.}, 
\href{http://arxiv.org/pdf/cond-mat/9804015v1}
{arXiv:cond-mat/9804015 .}

\bibitem{Tim99} E. Timmermans, P. Tommasini, R. C\^{o}t\'{e}, M. Hussein, A. Kerman, 
\href{http://dx.doi.org/10.1103/PhysRevLett.83.2691}
{Phys. Rev. Lett. {\bf 83}, 2691 (1999).}

\bibitem{Tim01} {E. Timmermans {\sl et al.}, }
\href{http://dx.doi.org/10.1016/S0375-9601(01)00346-2}
{Phys. Lett. A {\bf 285}, 228 (2001).}

\bibitem{Hol01} M. Holland, S.J.J.M.F. Kokkelmans, M.L. Chiofalo, and R. Walser, 
\href{http://link.aps.org/doi/10.1103/PhysRevLett.87.120406}
{Phys. Rev. Lett. {\bf 87}, 120406 (2001)}.

\bibitem{Kok02} {S.J.J.M.F. Kokkelmans {\sl et al.}, }
\href{http://link.aps.org/doi/10.1103/PhysRevA.65.053617}
{Phys. Rev. A {\bf 65}, 053617 (2002).}

\bibitem{Pri11c} L. Pricoupenko, M. Jona-Lasinio, 
\href{http://dx.doi.org/10.1103/PhysRevA.84.062712}
{Phys. Rev. A {\bf 84}, 062712 (2011).}

\bibitem{UVlimit} For ${\epsilon=0}$, this behavior gives the  ${1/r_{ij}}$ singularity of the 
wave function of an interacting pair $(ij)$ in the limit of a short interparticle distance 
${r_{ij}}$.

\bibitem{unexpected} The term in the second line of Eq.~\eqref{eq:STM_many_bosons} is in general 
not equal to zero when configurations with more than one molecule are possible. It was already 
found in the four-body problem if one performs the appropriate limit in Eq.~(50) of 
Ref.~\cite{Mor11a}. 

\bibitem{Mor11a} C. Mora, L. Pricoupenko, and Y. Castin, 
\href{http://dx.doi.org/10.1016/j.crhy.2010.12.005}
{C. R. Physique {\bf 12}, 71 (2011).}

\bibitem{one-mol-state} For the states where ${|\psi^m_\epsilon\rangle=0}$ for ${m \ge 2}$, in 
the limit of an arbitrarily large wave number ${k_n}$ (${n\ne 1,2}$), the function 
${\langle (k)_N^{\hat{1}\hat{n}}|D \rangle}$ is the only nonvanishing term in the left hand 
side of Eq.~\eqref{eq:STM_many_bosons}. Hence, in this limit ${\langle (k)_N^{\hat{1}\hat{2}} | D \rangle \sim 4\int d^3k_2 \langle (k)_N^{\hat{1}\hat{n}} 
| D \rangle/(3\pi^2 R^\star k_n^4)}$. Using this result and considering the limit ${k_{ij} \to \infty}$ 
for ${\{i,j\}\notin\{1,2\}}$, all the terms in the left hand side of Eq.~\eqref{eq:STM_many_bosons} 
vanish except the term with ${\langle (k)_N^{\hat{i}\hat{j}}|D \rangle}$. Thus in this limit 
${\langle (k)_N^{\hat{1}\hat{2}} | D \rangle \sim  
\int d^3k_2 \langle (k)_N^{\hat{i}\hat{j}} | D \rangle/(2\pi^2{R^\star k_{ij}^4})}$. The other solutions of 
Eq.~\eqref{eq:STM_many_bosons} containing more than one molecule follow a ${1/k_{ij}^2}$ 
law for ${k_{ij}\to \infty}$.

\bibitem{Pri08} L. Pricoupenko, 
\href{http://dx.doi.org/10.1103/PhysRevLett.100.170404}
{Phys. Rev. Lett. {\bf 100}, 170404 (2008).}

\bibitem{Pri11a} L. Pricoupenko, 
\href{http://dx.doi.org/10.1103/PhysRevA.83.062711}
{Phys. Rev. A {\bf 83}, 062711 (2011).}

\bibitem{Identity} In the limit where ${\epsilon \to 0^{+}}$, the following identity is used 
${
\int d^3k \chi_\epsilon^2(k)/(b^2+k^2)= (2\pi)^{3/2}(\epsilon^{-1}-b\sqrt{\pi/2} + b^2\epsilon) + \dots
}$.

\bibitem{Pri06a} L. Pricoupenko, 
\href{http://dx.doi.org/10.1103/PhysRevA.73.012701}
{Phys. Rev. A {\bf 73}, 012701 (2006).}

\bibitem{Pri11d} L. Pricoupenko, 
\href{http://arxiv.org/pdf/1110.5139}
{arXiv:1110.5139 .}

\bibitem{independent} This state does not depend on 
${\epsilon_{ij}, \dots \epsilon_{ln},\epsilon_{pq}}$.

\bibitem{Identity2} The identity ${N(N-1)P_{m-1}^{N-2}=2mP_m^N}$ is used.

\bibitem{Wer09} F. Werner, L. Tarruell, and Y. Castin, 
\href{http://www.springerlink.com/index/84j378181w489491.pdf}
{Eur. Phys. J. B {\bf 68}, 401 (2009).}


\bibitem{Wer12ab} F. Werner and Y. Castin, 
\href{http://dx.doi.org/10.1103/PhysRevA.86.013626}
{Phys. Rev. A {\bf 86}, 013626 (2012);}  
\href{http://dx.doi.org/10.1103/PhysRevA.86.053633}
{Phys. Rev. A {\bf 86}, 053633  (2012).}



\bibitem{Chi10} C. Chin {\sl et al.}, 
\href{http://link.aps.org/doi/10.1103/RevModPhys.82.1225}
{Rev. Mod. Phys. {\bf 82}, 1225 (2010).}


\bibitem{Pas12} B. Pasquiou, E. Mar\'echal, L. Vernac, O. Gorceix, B. Laburthe-Tolra,
\href{http://link.aps.org/doi/10.1103/PhysRevLett.108.045307}
{Phys. Rev. Lett. {\bf 108}, 045307 (2012).}

\end{thebibliography}
\end{document}